\documentclass[aps,prd,twocolumn,superscriptaddress,nofootinbib,eqsecnumm,showpacs]{revtex4-1}
\usepackage[pass,letterpaper]{geometry}
\pdfoutput=1  
\usepackage[]{graphicx}
\usepackage{color}
\usepackage{latexsym,amsmath,amsfonts,amssymb}
\usepackage[pdfencoding=auto]{hyperref}
\usepackage{url}
\usepackage{placeins}
\usepackage{hhline}
\usepackage{cancel}
\usepackage{booktabs}
\usepackage{verbatim}
\definecolor{MyBlue}{rgb}{0.15,0.15,0.70}
\hypersetup{
colorlinks=true,
citecolor=MyBlue,
linkcolor=MyBlue,
urlcolor=
}

\newcommand{\be}{\begin{equation}}
\newcommand{\ee}{\end{equation}}
\newcommand{\beq}{\begin{equation}}
\newcommand{\eeq}{\end{equation}}
\newcommand{\bea}{\begin{eqnarray}}
\newcommand{\eea}{\end{eqnarray}}

\newcommand\ees{\end{eqnarray}}
\newcommand\bees{\begin{eqnarray}}

\begin{document}

\title{Constraints on chameleon gravity from the measurement of the electrostatic stiffness of the MICROSCOPE mission accelerometers}

\author{Martin Pernot-Borr\`as}
\email{martin.pernot\_borras@onera.fr}
\affiliation{DPHY, ONERA, Universit\'e Paris Saclay, F-92322 Ch\^atillon, France}
\affiliation{Institut d'Astrophysique de Paris, CNRS UMR 7095,
Universit\'e Pierre \& Marie Curie - Paris VI, 98 bis Bd Arago, 75014 Paris, France}

\author{Joel Berg\'e}
\email{joel.berge@onera.fr}
\affiliation{DPHY, ONERA, Universit\'e Paris Saclay, F-92322 Ch\^atillon, France}

\author{Philippe~Brax}
\affiliation{Institut de Physique Th\'eorique, Universit\'e Paris-Saclay, CEA, CNRS, F-91191 Gif-sur-Yvette Cedex, France}

\author{Jean-Philippe Uzan}
\email{uzan@iap.fr}	
\affiliation{Institut d'Astrophysique de Paris, CNRS UMR 7095,
Universit\'e Pierre \& Marie Curie - Paris VI, 98 bis Bd Arago, 75014 Paris, France}
\affiliation{Sorbonne Universit\'es, Institut Lagrange de Paris, 98 bis, Bd Arago, 75014 Paris, France}

\author{Gilles M\'etris}
\affiliation{Universit\'e C\^ote d'Azur, Observatoire de la C\^ote d'Azur, CNRS, IRD, G\'eoazur, 250 avenue Albert Einstein, F-06560 Valbonne, France}

\author{Manuel Rodrigues}
\affiliation{DPHY, ONERA, Universit\'e Paris Saclay, F-92322 Ch\^atillon, France}

\author{Pierre Touboul}
\affiliation{DPHY, ONERA, Universit\'e Paris Saclay, F-92322 Ch\^atillon, France}

\date{\today}
\begin{abstract}
This article is dedicated to the use the MICROSCOPE mission's data to test chameleon theory of gravity. We take advantage of the technical sessions aimed to characterize the electrostatic stiffness of MICROSCOPE's instrument intrinsic to its capacitive measurement system. Any discrepancy between the expected and measured stiffness may result from unaccounted-for contributors, i.e. extra-forces. This work considers the case of chameleon gravity as a possible contributor. It was previously shown that in situations similar to these measurement sessions, a chameleon fifth force appears and acts as a stiffness for small displacements. The magnitude of this new component of the stiffness is computed over the chameleon's parameter space. It  allows us to derive constraints by excluding any force inconsistent with the MICROSCOPE data. As expected --since MICROSCOPE was not designed for the purpose of such an analysis--, these new bounds are not competitive with state-of-the-art constraints, but they could be improved by a better estimation of all effects at play in these sessions. Hence our work  illustrates this novel technique as a new way of constraining fifth forces.
\end{abstract}

\maketitle

\section{Introduction}

This article follows up from a series of articles~\cite{PRD1, PRD2,PRD3} aiming to test modified gravity theories with data from the MICROSCOPE mission. This mission provided the tightest constraint to date on  the weak equivalence principle (WEP) \cite{touboul_microscope_2017, Touboul_2019}. Its instrument is based on a couple of accelerometers measuring  the differential acceleration of two cylindrical test masses of different compositions. It contains four test masses: two cylinders of different composition in the SUEP (Equivalence Principle test Sensor Unit) sensor unit that is used to perform the WEP test and two cylinders of same composition in the SUREF (Reference Sensor Unit) sensor unit used as a reference. In Ref.~\cite{PRL}, we directly used the WEP test results to improve the current constraints on the existence of unscreened scalar fifth forces, a massive Yukawa fifth force and a light dilaton field \cite{PhysRevD.82.084033}.

In Ref.~\cite{CQG2}, we proposed a new way of testing such theories by using sessions dedicated to measuring the electrostatic stiffness inherent to the capacitive measurement system of MICROSCOPE. An electrostatic destabilizing force appears when a test mass is displaced from its rest position: it is linearly dependent to this displacement in the limit where it is small. We call {\em stiffness} its associated linear factor. It has been measured by applying a sinusoidal displacement of each test mass separately with an amplitude of $5\,\mu{\rm m}$.  The result of this series of tests has been compared to electrostatic models and a discrepancy has been pinpointed \cite{Chhun}. In Ref.~\cite{CQG2}, we modeled the total stiffness and studied all possible sources of forces to explain this discrepancy. They consist of mainly: (1) the satellite Newtonian self-gravity and (2) the stiffness of a 7-$\mu{\rm m}$-thick-gold-wire used to control the electrical potential of the test masses that acts as a spring. We found that the contribution of the former is sub-dominant. After determining the parameters of the latter to evaluate its contribution to the stiffness, we found an unexplained residual component that depends on the electrical configuration, hinting at patch field effects. We nonetheless considered the possibility that this discrepancy may originate from modified gravity fifth forces sourced by the satellite and experimental apparatus. We have already been able to set constraints on a Yukawa-like interaction by excluding any parameters of the interaction that lead to a stiffness larger than the discrepancy~\cite{CQG2}. As expected, since MICROSCOPE was not originally designed to such a test --leading to a loose estimation of the gold-wire-stiffness for instance--, the constraints are not competitive with state-of-the-art constraints but it opens a possible novel way of testing fifth force and demonstrate that its effect has to be modeled in details at each step of the experiment.

This article aims to extend this analysis to the chameleon gravity model \citep{khoury_chameleon_2004a,khoury_chameleon_2004}. Unlike Yukawa model, this scalar field enjoys a screening mechanism that makes its fifth force more sensitive to the matter environment and more subtle to compute. We use the numerical methods developed in Refs.~\cite{PRD1, PRD2} to compute the chameleon profile associated to a geometry of nested cylinders. In these articles, we studied the case of dis-centering one of the cylinders and showed that it should experience a chameleonic force acting as a stiffness for small displacements. Its magnitude depends on the geometrical parameters of the cylinders and on the parameters of the chameleon theory. This study was performed for only two nested cylinders. Here, we extend this methods to compute the field and the force associated to the geometry of the MICROSCOPE's instrument with the proper geometrical parameters. Each sensor unit is composed of eight cylinders: two cylindrical test mass cylinders, each of which is surrounded by two electrode cylinders; and two ferrule cylinders encompassing all the six cylinders \cite{Touboul_2019}. The end of these ferrules are closed by two ``lids" that we do not consider in this study.

This article is organized as follows. In Section II, we detail the methods used to compute the chameleon stiffness, and more particularly the necessity of different approximations for the different regimes of the chameleon gravity. In Section III, we present the constraints obtained by combining the results of these computations and the analysis of the MICROSCOPE stiffness measurement sessions from Ref.~\cite{CQG2}. To finish, in Section IV, we discuss our results and the limits of this new approach.

\section{Chamelon stiffness}

\subsection{Methods}
We use three different methods to compute the chameleon stiffness depending on the regimes of the chameleon field. These regimes occur for the MICROSCOPE geometry for different zones of the chameleon parameter space \cite{PRD1}. The chameleon field is parameterized by three parameters: its coupling constant to matter $\beta$, and the energy scale $\Lambda$ and index $n$ of its inverse-power law potential. We can distinguish three main regimes: the {\em screened regime} in which a test mass and the two electrode cylinders surrounding it can be considered as an isolated system due to the fact that the electrode cylinders screen the field; a {\em deeply screened regime} in which the screening of the test mass is too deep to compute the profile associated to three cylinders and instead we need to consider it as two separate pairs of screened cylinders; and a {\em unscreened regime} in which the field penetrates all cylinders so that all of them must be taken into account when computing the field profile.  Let us detail the computation techniques used in each regime.

\paragraph{Screened regime.}
This regime appears when the Compton wavelengths of the field in the cylinders are of the order of a twentieth of their thickness. It can be addressed by using the semi-analytic 2D model we developed in Ref.~\cite{PRD2}. This method was initially applied to two cylinders. Here, we modify it to include a third one. We impose the boundary conditions in the two external cylinders in such a way that the field must reach the minimum of the potentials associated to their densities. We displace the central test mass cylinder and solve the field's multipole from which we compute the force.

\paragraph{Deeply screened regime.}
This regime occurs when the Compton wavelengths are smaller than a twentieth of the cylinder's thickness. In this regime the screening of the test mass makes it impossible to use the previous method, as the value of the field reached deep in the test mass is so close to the value that minimizes its potential, that it is smaller than the typical numerical precision of a computer. We instead use a 1D method, and consider the three cylinders as two distinct pairs of screened parallel walls. To mimic two opposite sides of the cylinders, we consider two such systems. This 1D approximation is justified by the fact that we showed, in Ref.~\cite{PRD2}, that the chameleonic force computed in these planar and cylindrical configurations lead to the same order of magnitude for the acceleration experienced by a test mass. We thus postulate, for these analogous situations, that the test masses' accelerations verify $a_{\rm 2D} = \alpha \, a_{\rm 1D}$, where $\alpha$ is a geometrical factor that is expected to be of order unity.

\begin{figure}
\includegraphics[width = \columnwidth]{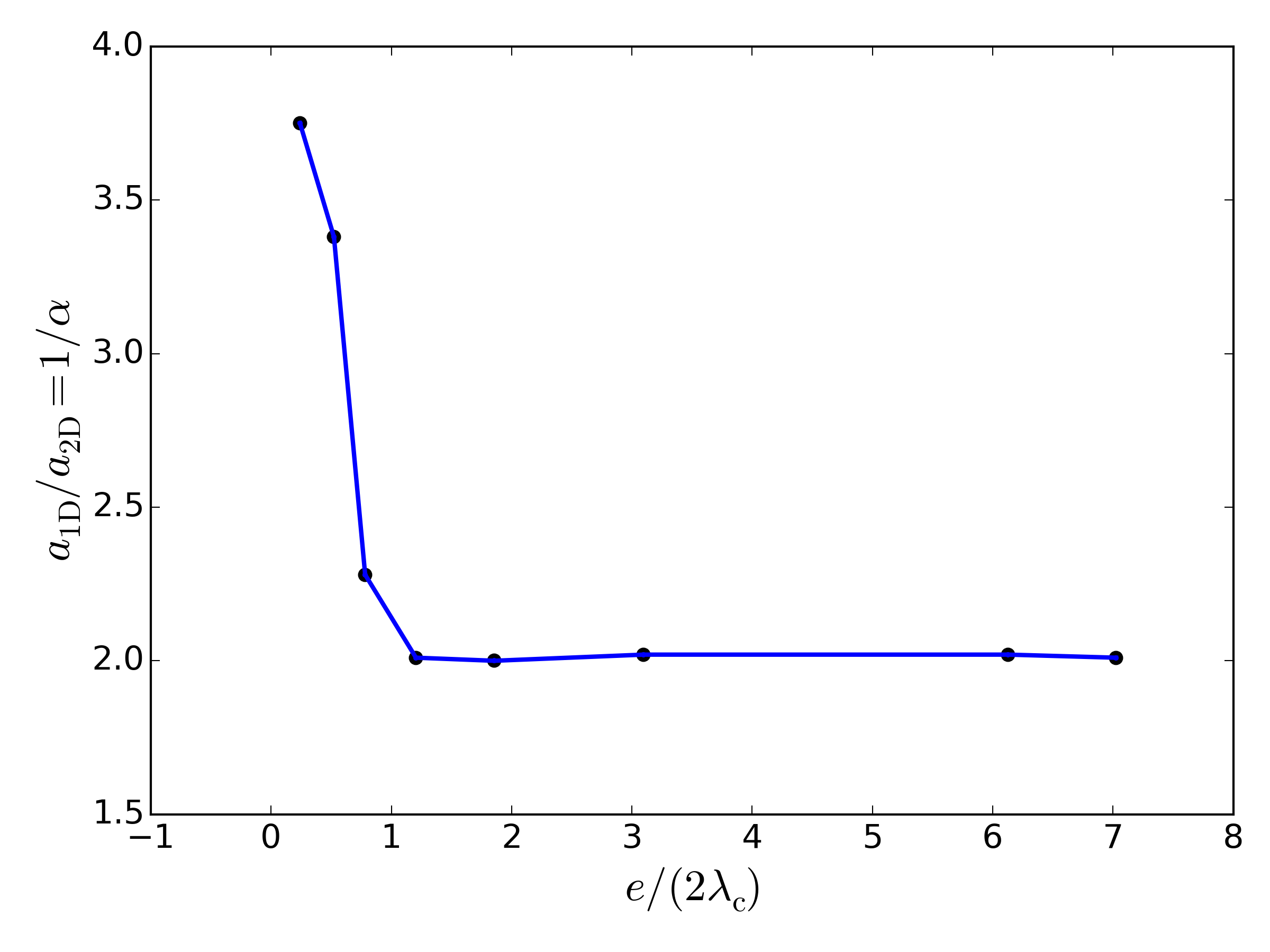}
\caption{Result of the comparison of the acceleration experienced by a test mass in planar and cylindrical configurations in different regimes of screening quantified by $\frac{e}{2 \lambda_{\rm c}}$, where $e$ is the thickness of the cylinder and $\lambda_{\rm c}$ the Compton wavelength of the field associated to it.}
\label{fig:alpha}
\end{figure}

From this equality, by using Newton's law, one can obtain a relation between the surface force $F_{\rm s, 1D}$ experienced by the two walls in a planar configuration and the force per unit length $F_{\rm l, 2D}$ experienced by a cylinder in the corresponding 2D configuration. The ratio of masses leads to the ratio of the wall thicknesses and the transverse section area of the cylinder in the relation
\begin{equation}
F_{\rm l, 2D} \approx \alpha \frac{\pi \left[(d+ e)^2 - d^2\right]}{2\, e} F_{\rm s, 1D},
\label{eq:F_equiv}
\end{equation}
where $d$ and $e$ are respectively the internal radius and the thickness of the test mass cylinder. The value of $\alpha$ is discussed in Fig.~\ref{fig:alpha} and below.

\paragraph{Unscreened regime.}
This regime takes place when the Compton wavelengths of the field in the cylinders are larger than their thicknesses. In this case the boundary conditions must be set at some distance much larger than the Compton wavelength associated to the density outside the cylinders. In this regime, this Compton wavelength is likely to be so large that one must perform large steps in terms of the numerical resolution in this zone, hence losing the accuracy on the result. To overcome this issue we again addressed this regime with a 1D resolution. In a 1D problem, as discussed in Ref.~\cite{PRD1}, the chameleon equation can indeed be integrated once in the region external to the cylinders and obtain, at the boundary of the external cylinder, a condition $\phi'[\phi(x_{\rm b})]$ giving the field derivative as a function of the field value, which ensure that the boundary conditions are respected far from it.  We can use this condition to perform a dichotomy method to adjust the initial condition of our numerical method. We proceed in the same way as the case of asymmetrical parallel walls in Ref.~\cite{PRD2}, with the difference that instead of using for the dichotomy method the verification that the boundary conditions are respected at some large distance from the cylinders, we check that the aforementioned condition is verified at the boundary of the outer cylinder.  These two conditions are equivalent but the latter allows us to bypass  solving the field in the external region.

Similarly to the previous regime, in this 1D resolution, to mimic two opposite radial directions of a 8-nested-cylinder-configuration, we consider a set of 16 parallel walls. In this 1D configuration a test mass is represented by two of these walls. We again use Eq.~\eqref{eq:F_equiv} to compute the corresponding 2D force. Note that due to the symmetry breaking by the shifting of the walls, the initial conditions cannot be set at the center of the 16 walls but instead at a slightly shifted location that we determine similarly as in Ref.~\cite{PRD2}.

\vspace{.5cm}

To evaluate $\alpha$ in Eq.~\eqref{eq:F_equiv}, we compare the forces computed in 1D and 2D. This requires to extend the method used in the screened regime to the other regimes. To overcome the problem encountered in these regimes, we considered an unrealistic configuration of 3 cylinders of same density with a external vacuum much denser than the vacuum of space. This allows us to avoid the numerical resolution issue encountered in the unscreened regime. Even if unrealistic, it allows us to quantify the geometrical factor between planar and cylindrical geometries, that we expect to be independent of the densities.

As depicted in Fig.\ref{fig:alpha}, the numerical comparison strongly hints at $\alpha
= 1/2$, a value reached in most of the screening range but that appears to be smaller for unscreened situations. We interpret this latter behavior as the 2D method reaching its limits and we instead expect $\alpha = 1/2$ also this regime. This is justified by the longer Compton wavelength in this regime, that leads the field's
gradient to vary slowly within the cylinder. By approximating this gradient by the one obtained in planar situations, one directly obtains Eq.\eqref{eq:F_equiv} with $\alpha= 1/2$ \footnote{The origin of this value comes from the fact that while for planar situations all parts of a wall are subjected to a force, for cylindrical configurations, only the parts of the cylinder that are closer to the axis of displacement contribute to the acceleration. This is due to the projection of the force that is mainly radially directed and to the effective radial displacement of the cylinder that varies with the cylindrical angle.}. Hence we choose to generalize this result to all regimes in our present study.

\subsection{Results}
\begin{figure}
\includegraphics[width = \columnwidth]{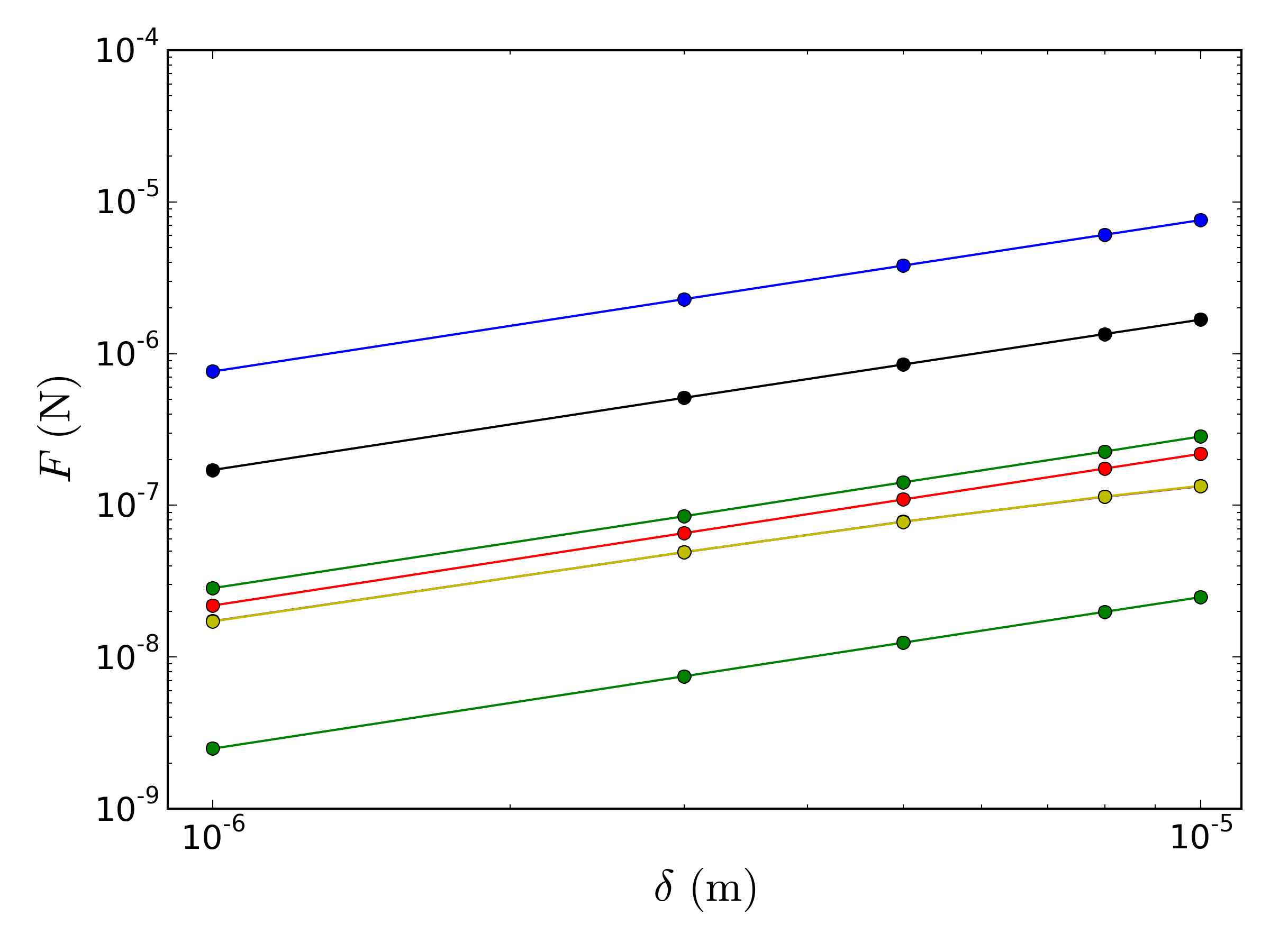}
\caption{Scaling of the chameleon fifth force as a function of the displacement in the range $\delta=1\ldots 10~\mu$m for different set of parameters ($\Lambda,\beta)$ assuming $n=1$. The force is expressed in newtons. $\Lambda$ are chosen in the range $10^{-1}-3\times10^2$~eV and $\beta$ in the range $6-10^7$. This shows that $\log F=\log k_{\rm chameleon}(\Lambda,\beta) + \log \delta$ is a good approximation to the behavior of the force at small displacements. We use a log-log plot for convenience but it is easily checked that the slope is unity so that linearity is confirmed.}
\label{fig:linear}
\end{figure}	

First, we check numerically that the force is linear for small displacements. As shown in Ref.~\cite{PRD2}, this is expected to be the case even though the theory is non-linear. Figure~\ref{fig:linear} depicts the behavior of $F(\delta)$ in the range $\delta=1\ldots 10~\mu$m relevant for our study. Besides, we know that by symmetry $F(0)=0$. Hence it confirms that in this range of displacements it is safe to model the chameleon fifth force by a stiffness $k_{\rm chameleon}(\Lambda,\beta)$ (measured in N.m$^{-1}$) so that
\begin{equation}
 F = k_{\rm chameleon}(\Lambda,\beta) \times \delta +{\cal O}(\delta^2).
\end{equation}
Even though one can witness a small deviation of this linear relation for $\delta\sim 10~\mu$m for the largest values of $\Lambda$, these results comfort us in the linearity assumption in the range of displacements compatible with the MICROSCOPE data we are using and the parameter space we consider.

\begin{figure}
\includegraphics[width = \columnwidth]{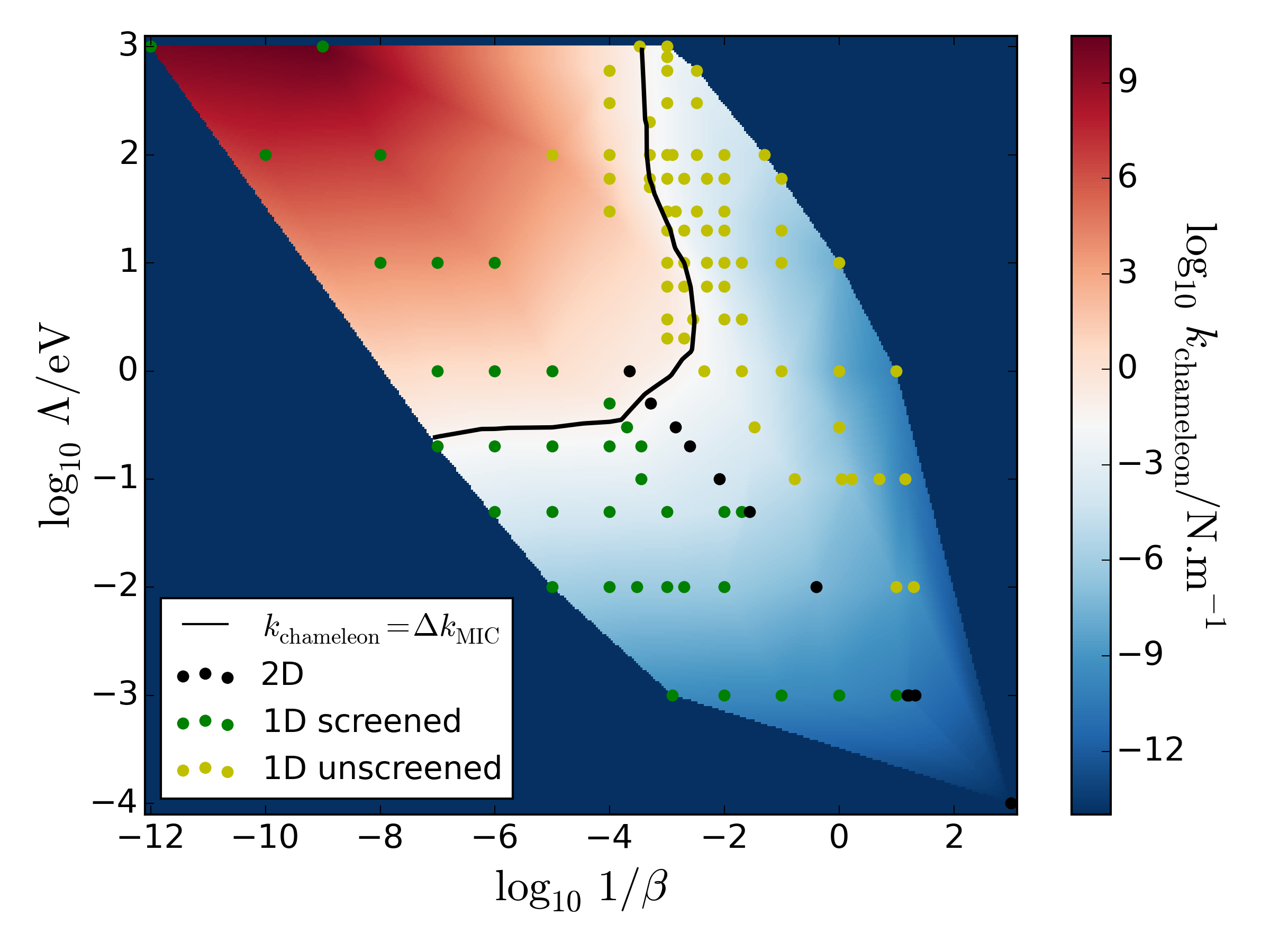}
\caption{Evolution with the parameters $\beta$ and $\Lambda$ of the chameleonic stiffness $k_{\rm chameleon}$ of the external test mass of the sensor unit SUEP from the MICROSCOPE mission for $n = 1$. Its magnitude is shown by the background colors. This function is obtained by linearly interpolating the data points. These points are the result of the three numerical methods discussed in the main text that are here distinguished by different points of colors. This is represented in log-scale for $\beta$, $\lambda$, $k_{\rm chameleon}$. The uniform-dark-blue-region corresponds to parameters for which we are unable to compute the stiffness. The black line is the contour line at which the stiffness is equal to the measured 2-$\sigma$ uncertainty on the discrepancy $\Delta k_{\rm MIC}$ in the MICROSCOPE experiment.}
\label{fig:kpoints}
\end{figure}

Then, we present the numerical results in Fig.~\ref{fig:kpoints}. We computed the chameleonic stiffness $ k_{\rm chameleon}(\Lambda,\beta)$ experienced by a test mass when displaced radially by $1 \,\mu{\rm m}$. This figure shows the result for SUEP-IS2, the external test mass of SUEP. We spanned the parameter space $(\beta,\Lambda)$ for $n = 1$, we denote each computation by a point with a color code that labels which of the three method were used. To obtain the continuous evolution of the stiffness with $(\beta,\Lambda)$, we performed a linear interpolation of the simulation points in log-scale. We show with the black solid line, the contour line at which the obtained chameleonic stiffness equals the 2-$\sigma$ uncertainty on the discrepancy $\Delta k_{\rm MIC}$ on the stiffness measured in the MICROSCOPE sessions as presented in Ref.~\cite{CQG2}. This latter article presents two distinct estimations over two perpendicular radial axis of the cylinder; the chameleonic stiffness being expected to be the same over these axes, we choose to average these two estimations and quadratically average the error bars. 

\section{Constraints on the chameleon's parameters}

The results shown in Fig.~\ref{fig:kpoints} mean that above the black line the chameleonic stiffness is too large to explain the observed stiffness residual in MICROSCOPE. This stiffness could be compatible with these measurements, if a stabilizing stiffness of the same magnitude were to exist. Nevertheless standard physics combined with our understanding of the instrument do not provide any such contribution. Thence we interpret these results as excluding the existence of a chameleon field for these parameters. Below the black line, the chameleonic stiffness is within the error bars of the observed discrepancy so that we cannot exclude its existence.

Note that we have not been able to span the whole parameter space. Our methods are unable to determine the stiffness for large $\beta$ and $\Lambda$. We expect this to be caused by the fact that the field magnitude becomes so large that our numerical precision fails at describing the gradient in the test mass. Thus, the force vanishes. Nevertheless we can guess the behavior of the stiffness in these unexplored regions. For very large $\Lambda$, the field tends to be completely unscreened such that we expect it to converge towards a flat field providing a lower force. For very large $\beta$, on the contrary, the field tends to be more screened. At some point we expect the field to be able to reach the minimum of its potential in the inter-cylinder vacuum gaps, such that the cylinders would not interact through the scalar field anymore. In this case the field is equivalent to the field of an infinitely thick cylinder and gap. Given the inter-cylinder gaps of $600\,\mu{\rm m}$ for MICROSCOPE, we expect this to happen for $\beta\gtrsim10^{19}$. We thus expect the MICROSCOPE constraint to have a rectangular shape.

\begin{figure}
\includegraphics[width = \columnwidth]{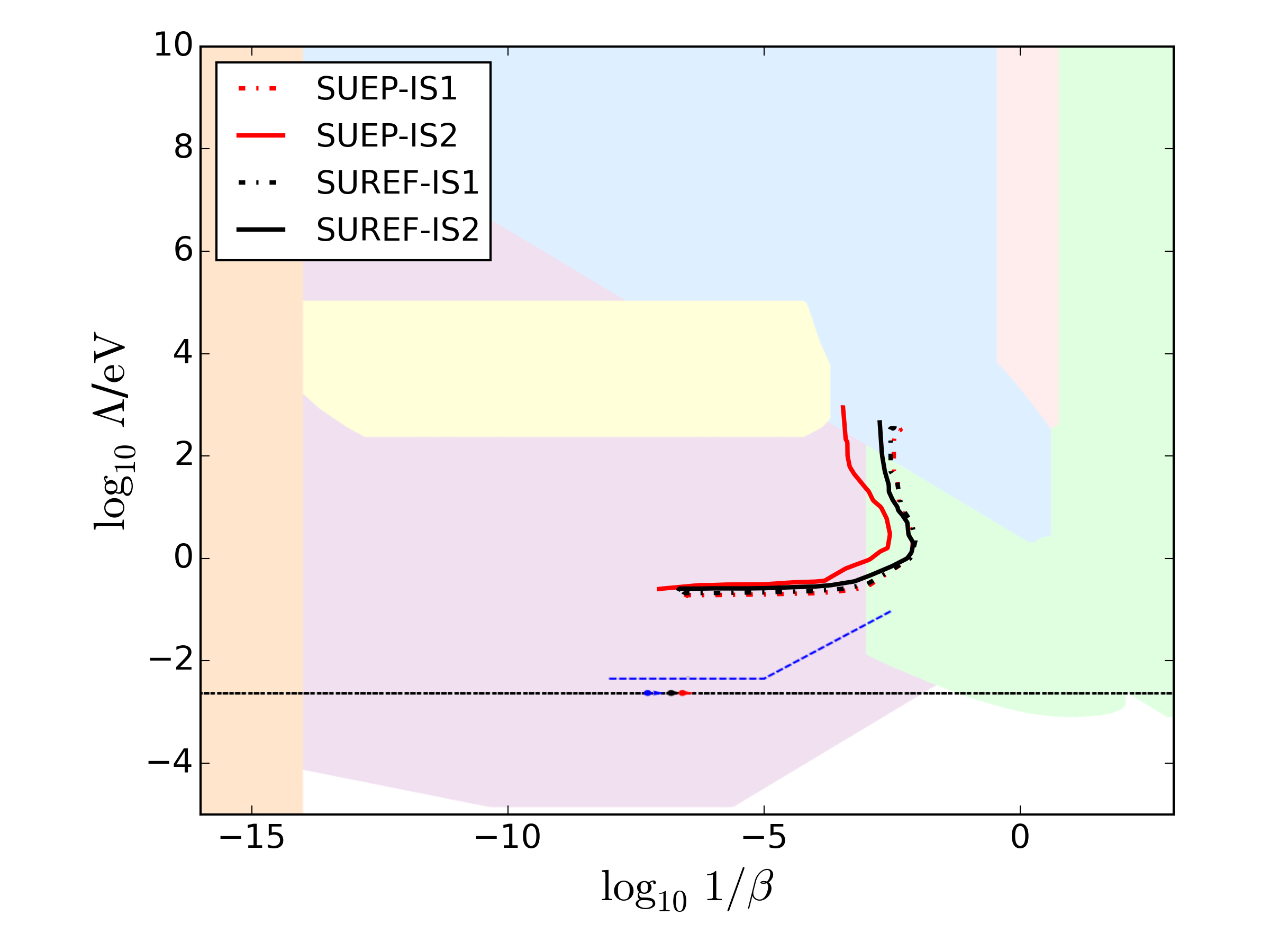}
\caption{Constraints on the chameleon model for $n = 1$ from the MICROSCOPE experiment using stiffness measurement sessions: the region excluded at 2-$\sigma$ is above the four lines described in the legend. They correspond to the different test masses: IS1 (resp. IS2) denotes the the internal (resp. external) test masses of the SUREF and SUEP sensor units. Their constraints are compared to the current constraints from other experiments denoted with the colored regions as presented in Refs.~\citep{burrage_tests_2018, BraxReview}. They come from atomic interferometry (purple, \cite{hamilton_atom-interferometry_2015,jaffe_testing_2017}), E\"ot-Wash group's torsion balance experiments (green, \citep{upadhye_dark_2012,PhysRevLett.98.021101}), Casimir effect measurements (yellow, \cite{brax_detecting_2007,PhysRevD.75.077101}), astrophysics tests (blue, \cite{Jain_2013,Cabr__2012,PhysRevD.97.104055}) and lensing (pink, \cite{10.1093/mnras/stw1617}), precision atomic tests (orange, \cite{PhysRevD.83.035020, PhysRevD.82.125020}), microsphere (blue line, \cite{PhysRevLett.117.101101}) and neutron interferometry (blue and red point, \cite{LEMMEL2015310,PhysRevD.93.062001}). The horizontal doted line denotes the energy scale of dark energy.}
\label{fig:MIConstr}
\end{figure}

We applied the same procedure to the other three test masses. The result are summarized in Fig.~\ref{fig:MIConstr}. It shows the 2$\sigma$-constraints from each test mass: the internal mass of each sensor unit is called IS1 and the external IS2. We compare the MICROSCOPE constraints to the current constrains summarized in Refs.\cite{burrage_tests_2018,BraxReview}. They overlap the constraints from atom interferometry \cite{hamilton_atom-interferometry_2015,jaffe_testing_2017}, torsion balances \citep{upadhye_dark_2012,PhysRevLett.98.021101} and Casimir effect experiments \cite{brax_detecting_2007,PhysRevD.75.077101}. Nevertheless, they are not competitive with current constraints. This is not surprising since MICROSCOPE was not designed for this test.

\section{Discussion}
The best constraints are obtained from the internal test masses --IS1. This is explained by a better estimation --by one order of magnitude-- of the gold-wire-stiffness \cite{CQG2} leading to a lower residual stiffness. The competitivity of the internal masses is nonetheless depleted by the shortness of these test masses relative to the external ones \cite{Touboul_2019}. We observe that the constraints from the internal test masses are very similar, their slight difference is only caused by a slightly different residual stiffness. They indeed experience the same chameleonic stiffness, which is consistent with the fact that they have the same geometrical parameters and are of the same composition. This tells us that the effect on the inner masses from the external test mass --of different compositions for the two sensor units-- is negligible even in the unscreened regime --upper part of the constraint.

Comparing the chameleonic forces of the external test masses --that have same geometrical parameters but different densities-- is interesting for the phenomenology of a WEP violation. This requires to normalize them by their masses. Doing so reveals that they each experience, in these dis-centered configurations, a different acceleration in both screened and unscreened regime. This confirms the ability of the chameleon field to provide an apparent WEP-violation-signal as the only result of the different densities of test masses through their different screening factors \cite{khoury_chameleon_2004}. This has no direct repercussion on MICROSCOPE's WEP test as: (1) it is performed in a situation where coaxiality of all cylinders is well controlled \cite{Touboul_2019}, (2) it is performed on a couple of test masses --IS1 and IS2-- belonging to the same sensor unit for which the different geometrical parameters could also be the source of a differential acceleration. This dependence of the force to the test masses' densities nonetheless hints at an apparent-chameleonic-WEP-violation to appear in MICROSCOPE's WEP test. Note that the common wisdom about chameleon inducing apparent WEP violation in screened regimes \citep{khoury_chameleon_2004a,khoury_chameleon_2004} is not applicable to MICROSCOPE's test of the WEP, since in this case, the satellite itself screens the Earth's chameleon field \cite{PRD1}, preventing any WEP violation signal at the frequency aimed by MICROSCOPE. Instead, we expect such a signal to appear in a lightly screened regime where the Earth's chameleon profile can penetrate the instrument. Of course in such a regime the density dependence of the force would be depleted but the signal it induces might still be detectable if the precision of the experiment is high enough. Estimating this effect is beyond the scope of this article.

We obtained these new constraints from numerical simulations of the chameleon profiles in the nested-cylinder-geometry of the MICROSCOPE experiment. Some approximations must be discussed. Firstly, when evaluating the chameleonic stiffness, we used the profiles of infinitely extended cylinders. In MICROSCOPE, the cylinders being finite, we expect edge effects to appear that would require 3D simulations to quantify and that are beyond the scope of this study. Nevertheless, we expect these effects to decrease the computed stiffness. We indeed predict the field to behave as follows. On the one hand far from the ends of a cylinder, the transverse profile should be close to the one of infinite cylinders. On the other hand, at its ends, it should be influenced by the two cylindrical ``lids" that close the ends of the electrode cylinders. We expect the presence of this matter to affect the chameleon profile in such a way that it is flattened in comparison to the profile of infinite cylinders. This flattening would reduce the gradients in the test mass at its ends, leading our computed stiffness to be overestimated. This would induce our constraints to be slightly decreased.

Another assumption is that we computed the profile for a static configuration while the stiffness measurement sessions involve a periodic motion of the test mass. The validity of this quasi-static assumption depends on the relaxation time of the field in response to a change in the matter distribution. We expect this assumption to stay valid as long as the movement are slow compared to the relaxation speed of the field. In analogy with gravitational waves \cite{PhysRevD.57.2061}, and consistently with the discussion from Ref.~\cite{burrage_probing_2015}, we expect this speed to be close to the speed of light for light fields and lower for massive fields. This assumption could thus be questionable for chameleon parameters providing the heaviest fields such as in the deeply screened regime. Nevertheless, this regime is not accessible to our methods.

Finally, we idealized the MICROSCOPE geometry by not taking into account the influence of MICROSCOPE's satellite but only the effect of the instrument. This is debatable in the regime where the field is unscreened. The complex geometry of the satellite could introduce peculiar effects on the chameleonic force. Nonetheless, given the null-effect on the internal test mass of the external ones, and the low factor of 100 between the mass of the cylinders and of the satellite, we expect the influence of mass distribution closest to the test masses, i.e. the electrodes cylinders, to be dominant. This has for instance been demonstrated for a Yukawa fifth force in Ref.~\cite{CQG2}.

To conclude, this work extends the search for new methods to test chameleon models in the laboratory~\cite{PRD3} or in space~\cite{Berge:2019zjj,Berge:2018htm}. Here we took advantage of MICROSCOPE's instrumental characterization measurements to draw constraints on the chameleon field. An unexplained discrepancy between the measured and expected electrostatic stiffness might hint at a non-zero chameleonic force. The constraints we obtained are not competitive with state-of-the-art constraints. This is not a surprise. MICROSCOPE was not designed for testing short-ranged modified gravity theories.  The main limitations of this test come from modeling uncertainties of the theoretical electrostatic stiffness and from the poor knowledge of the gold-wire characteristics. A better estimation of these physical parameters would reduce the error bars on the stiffness discrepancy. An alternative, under study for a next mission \cite{battelier2019exploring}, is to suppress this gold wire as done in LISA Pathfinder \cite{PhysRevLett.116.231101}. Besides, patch field effects may be the most likely phenomenon to explain the observed discrepancy on the measurement of the stiffness \cite{CQG2}. Estimating these effects would deplete this discrepancy and thus improve the sensitivity of the test. While awaiting these developments, the constraints we have provided are conservative.

\section*{Acknowledgment}
We thank the members of the MICROSCOPE Science Working Group for allowing us to start this project and encouraging us to pursue it. We acknowledge the financial support of CNES through the APR program (``GMscope+'' project). MPB is supported by a CNES/ONERA PhD grant. This work uses technical details of the T-SAGE instrument, installed on the CNES-ESA-ONERA-CNRS-OCA-DLR-ZARM MICROSCOPE mission. This work is supported in part by the EU Horizon 2020 research and innovation programme under the Marie-Sklodowska grant No. 690575. This article is based upon work related to the COST Action CA15117 (CANTATA) supported by COST (European Cooperation in Science and Technology).

\appendix

\bibliographystyle{ieeetr}
\bibliography{Bib}

\end{document}